\documentclass[12pt]{iopart}
\usepackage{psfig}  

\newcommand{\be}{\begin{equation}}
\newcommand{\ee}{\end{equation}}
\newcommand{\bea}{\begin{eqnarray}}
\newcommand{\eea}{\end{eqnarray}}

\newcommand{\rar}{\rightarrow}

\newcommand{\mb}{\mathbf}
\begin{document}

\title[Measuring complexity with zippers]{Measuring complexity with zippers}

\author{Andrea Baronchelli\dag, 
Emanuele Caglioti\ddag $\:$  and Vittorio Loreto\dag}

\address{\dag\ Physics Dept. and INFM-SMC {\em La Sapienza}
University, P.le A. Moro 2, 00185 Rome, ITALY}

\address{\ddag\ Mathematics Dept. {\em La Sapienza}
University, P.le A. Moro 2, 00185 Rome, ITALY}

\begin{abstract}
Physics concepts have often been borrowed and independently developed
by other fields of science.  In this perspective a significant example
is that of entropy in Information Theory. The aim of this paper is to
provide a short and pedagogical introduction to the use of data
compression techniques for the estimate of entropy and other relevant
quantities in Information Theory and Algorithmic Information
Theory. We consider in particular the LZ77 algorithm as case study and
discuss how a zipper can be used for information extraction.
\end{abstract}




\section{Introduction}

Strings of symbols are nowadays widespread in all the fields of
science. On the one hand many systems are intrinsically described by
sequences of characters: DNA, written texts, bits in the transmission
of digital data, magnetic domains in storage data devices, etc. On the
other hand a string of characters is often the only possible
description of a natural phenomenon. In many experiments, for example,
one is interested in recording the variation in time of a given
physical observable (for instance the temperature of a system), thus
obtaining a time series, which, suitably codified, results in a
sequence of symbols.

Given a string of symbols the main problem is quantifying and then
extracting the information it contains. This acquires different
meanings in different contexts. For a DNA string, for instance, one
could be interested in separating portions coding for proteins from
not coding parts. Differently in a written text important information
are the language in which it is written, its author, the subject
treated etc.

Information Theory (IT) is the branch of science which deals, among
other things, with the problems we have mentioned. In a seminal paper
dated 1948 Claude Shannon pointed out the possibility of quantifying
the information contained in a (infinite) string of characters
\cite{shannon}. Adopting a probabilistic approach, i.e. focusing the
attention on the source generating a string, the famous
Shannon-McMillan theorem shows that there is a limit to the
possibility of compressing a string without loosing the information it
brings. This limit is proportional to the {\em entropy} (or informatic
content) of that string \cite{shannon,K57}. 

A remark is interesting now. The name entropy is not accidental, and
information theory represents one of the best examples of a concept
developed in physics whose role became of primary importance also in
another field. Historically, the concept of entropy was initially
introduced in thermodynamics in a phenomenological context. Later,
mainly by the contribution of Boltzmann, a probabilistic
interpretation of the entropy was developed in order to clarify its
deep relation with the microscopic structure underlying the
macroscopic bodies.  On his hand, Shannon, generalizing the concept of
entropy in the apparently unrelated field of communication systems,
was able to establish a self consistent information theory.  For a
recent excursus about the notion of entropy see~\cite{legacy}. We
shall describe more precisely Shannon's approach in the following
section, but we refer the interested reader to \cite{parisi} for a
discussion of the connections between Shannon and microscopic
entropies.

A radically different approach to the information problem,
namely the Algorithmic Information Theory (AIT)
~\cite{Ch66,Ch90,k65,S64}, was developed towards the half of the
1960s. It showed again, from a different point of view, that a good way of
quantifying the information embedded in a string is that of trying to describe
it in the shortest possible way.

In this framework it seems natural to look at those algorithms
expressly conceived to compress a file (i.e. a string of bytes), known
as {\em zippers}. A zipper takes a file and tries to minimize its
length. However, as we have mentioned, there is a theoretical limit,
represented by the entropy of the considered sequence, to the
performance of a zipper. A compression algorithm able to reach this
theoretical limit is said to be ``optimal''. Thus an optimal zipper
can be seen as an ideal tool to estimate the informatic content of a
string, i.e. to quantify the information it brings. In this paper we
shall discuss this possible application of data compression algorithms
together with its shortcomings.

Finally, besides the important scientific problem of measuring how
much information is contained in a string, one could ask if it is
possible to extract that information. With a slight abuse of the word,
we can address the level of the kind of information contained in a
sequence as the semantic level. We are then interested in asking
whether it is possible to access to the semantic level from a
information theoretical, ``syntactic'', analysis of a string. We shall
show that, under certain assumptions, this is indeed the case in many
different circumstances.

The outline of this paper is as follows. In Section II we make a short
introduction to some Information Theory concepts; in Section III we
describe the optimal compression algorithm LZ77; in Section IV,
finally, we illustrate with some examples the possible applications of
the illustrated information extraction techniques.


\section{Entropy and complexity}

In Shannon's probabilistic approach to information, born in an
engineering context, the communication scheme is fundamental.  A
message is first produced by a source of information, then is codified
in a way proper for the transmission in a channel and finally, before
arriving to the receiver, it must be brought back to the original
form.  \\ All these steps are of great theoretical interest, but for
our purposes we will concentrate on the source uniquely.  This is a
device able to form a message adding one symbol per unit time, chosen
in agreement with some probabilistic rules, to the previously emitted
ones. Here we consider only cases in which the possible characters are
finite in number, i.e. the alphabet $\mathcal{X}$ is finite. The
source can then be identified with the stochastic process it obeys.
Shannon's IT always deals with ergodic sources. A rigorous definition
of ergodic processes is out of the scope of this paper. We shall limit
ourselves to an intuitive definition. A source is ergodic if it is
stationary (the probability rules of the source do not vary in time)
and it holds the following property. If $N_l$ is the number of
occurrences of a generic sequence $Y=y_1, ...,y_s$ in a string $X$ of
length $l>s$, then:

\be \lim_{l \rightarrow \infty}
P\{|\frac{N_l}{l}-P(x_{i_1},...,x_{i_s} = y_1,...,y_s)|<\epsilon \} =
1 \;\;\;\;\;\; \forall \; \epsilon, y_s 
\ee

\noindent i.e. the averages made over an emitted string, $\frac{N_l}{l}$,
coincide with those made over time $P(x_{i_1},...,x_{x_s}= y_1, ...,
y_s)$, in the limit of infinite string length.

Now, if $\mb{x}$ is a $n$-symbols sequence chosen from the
$\mathcal{X}^n$ possible sequences of that length, we introduce the
$N-$block entropy as:

\be
H_n = H(X_1,X_2, ..,X_n)= -\sum_{\mb{x} \in
\mathcal{X}^n}p(\mb{x}) \log p(\mb{x})  
\ee

\noindent where $p(\mb{x})$ is the probability for the string $\mb{x}$
to be emitted. The differential entropy $h_n=H_{n+1}-H_n$ represents
the average information carried by the $n+1$ symbol when the $n$
previously emitted characters are known. Noting that the knowledge of
a longer past history cannot increase the uncertainty on the next
symbol, we have that $h_n$ cannot increase with $n$, i.e. it holds
$h_{n+1} \leq h_n$ and for an ergodic source we define the Shannon
entropy $h$ as:

\be
h = \lim_{n  \rar \infty} h_n = \lim_{n \rar \infty} \frac{H_n}{n}.
\ee

The entropy of a source is a measure of the information it
 produces. In other words $h$ can be viewed as a measure of the
 surprise we have analyzing a string generated by a stochastic
 process. Consider for example the case of a source emitting a unique
 symbol $A$ with probability 1. For that source it would hold $h=0$,
 and in fact we would have no surprise observing a new $A$. On the
 other hand if the probability of occurrence of the symbol $A$ is
 quite small our surprise will be proportionally large. In particular
 it turns out to be proportional to the absolute value of the
 logarithm of its probability. Then $h$ is precisely the average
 surprise obtained by the stochastic process. Remarkably it can be
 shown that $h$, apart from multiplicative coefficients, is the only
 quantity that measures the surprise generated by a stochastic
 process~\cite{K57}.  \\ More precisely, the role of $h$ as an
 information measure can be fully recognized in the Shannon-McMillan
 theorem \cite{shannon, K57}. Given a $N$ characters-long message
 emitted by an ergodic source, it states that:
 
\begin{enumerate}
\item It exists a coding for which the probability for the message to
  require more than $\; N h_2=(Nh/\log2) \;$ 
  bits tends to zero when $N$ tends to infinity.
\item It does not exist a coding for which the probability for the message
  to require less than $N h_2$ bits tends to one when $N$ tends to infinity.
\end{enumerate}
  
A completely different approach to information related problems is
that of the Algorithmic Information
Theory~\cite{Ch66,Ch90,k65,S64}. In this context the focus is on the
single sequence, rather than on its source, and the basic concept is
the Algorithmic Complexity: {\em the entropy of a string of characters
is the length (in bits) of the smallest program which produces as
output the string and stops afterwards}.  This definition is
abstract. In particular it is impossible, even in principle, to find
such a program and as a consequence the algorithmic complexity is a
non computable quantity.  This impossibility is related to the halting
problem and to Godel's theorem~\cite{livit}.  Nevertheless also this
second approach indicates that searching for the most concise
description of a sequence is the way for estimating the amount of
information it contains. As one could expect, in fact, there is a
connection between the Algorithmic Complexity of a string and the
Entropy of its source, but we refer the interested reader
to~\cite{livit} for a detailed discussion.

Up to this point our attention has been devoted to the
characterization of a single string. Both IT and AIT, however, provide
several measures of relations of remoteness, or proximity, between
different sequences. Among these, it is interesting to recall the
notion of relative entropy (or Kullback-Leibler
divergence)~\cite{KL,Kullback,cover-thomas} which is a measure of the
statistical remoteness between two distributions. Its essence can be
easily grasped with the following example.  \\ Let us consider two
stationary memoryless sources $\cal A$ and $\cal B$ emitting sequences
of $0$ and $1$: $\cal A$ emits a $0$ with probability $p$ and $1$ with
probability $1-p$ while $\cal B$ emits $0$ with probability $q$ and
$1$ with probability $1-q$. The optimal coding for a sequence emitted
by $\cal A$ codifies on average every character with $\; h({\cal{A}})
= -p \, \log_2 p -(1-p) \, \log_2(1-p) \;$ bits (the Shannon entropy
of the source). This optimal coding will not be the optimal one for
the sequence emitted by $\cal B$. In particular the entropy per
character of the sequence emitted by $\cal B$ in the coding optimal
for $\cal A$ will be:

\be C({\cal{A}|\cal{B}}) = -q \,\log_2 p - (1-q) \, \log_2 (1-p) 
\label{cross} \ee

\noindent while the entropy per character of the sequence emitted by
$\cal B$ in its optimal coding is $-q \, \log_2 q - (1-q) \, \log_2
(1-q)$.  Eq.(\ref{cross}) defines the so-called cross-entropy per
character of $\cal A$ and $\cal B$.  The number of bits per character
waisted to encode the sequence emitted by $\cal B$ with the coding
optimal for $\cal A$ is the relative entropy of $\cal A$ and $\cal B$:

\begin{equation}
d({\cal A}\vert \vert {\cal B})= C({\cal{A}|\cal{B}}) - h({\cal{A}}) =
-q \, \log_2 \frac{p}{q} - (1-q) \, \log_2
\frac{1-p}{1-q}.
\label{relative}
\end{equation}

\noindent A linguistic example will help to clarify the situation:
transmitting an Italian text with a Morse code optimized for English
will result in the need of transmitting an extra number of bits with
respect to another coding optimized for Italian: the difference is a
measure of the relative entropy between, in this case, Italian and
English (supposing the two texts are each one archetypal
representations of their Language, which is not).

It is important to remark that the relative and cross entropies are
not distances (metric) in the mathematical sense, since they are not
symmetric and do not satisfy in general the triangular
inequality. Defining a true distance between strings is an important
issue both for theoretical and practical reasons (see for some recent
approaches~\cite{bcl,otu-sayood,li_similarity} and for a short
review~\cite{long}).


\section{Zippers} 

In the previous section we have seen two different approaches to the
characterization of the information, the Classical and the Algorithmic
ITs. We have also seen that, despite their profound differences, both
of them indicate that the way to quantify the information of a string
is to find its shortest description, i.e. to compress it. Driven by
this fact, in this paragraph we shall illustrate the LZ77 compression
algorithm, that, asymptotically, is able to get to the Shannon limit.

\begin{figure}
\centerline{\psfig{figure=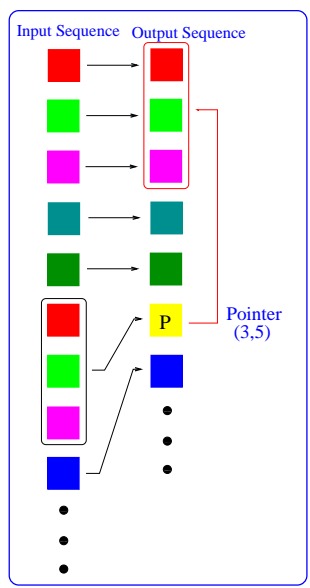,width=4cm,angle=0}}
\caption{{\bf Scheme of the LZ77 algorithm:} The LZ77 algorithm
searches in the look-ahead buffer for the longest substring (in this
case substring of colors) already occurred and replaces it with a
pointer represented by two numbers: the length of the matching and its
distance}
\label{fig0}
\end{figure}

The Lempel and Ziv algorithm LZ77~\cite{LZ77} (see Figure~\ref{fig0})
(used for instance by $gzip$ and $zip$ commercial zippers) achieves
compression exploiting the presence of duplicated strings in the input
data. The second occurrence of a string is replaced by a pointer to
the previous string given by two numbers: a distance, representing how
far back into the window the sequence starts, and a length,
representing the number of characters for which the sequence is
identical. More specifically the zipper reads sequentially the input
$N$-symbols sequence, $x=x_1,....,x_N$. When $n$ symbols have already
been analyzed, LZ77 finds the longest string starting at symbol $n+1$
which has already been encountered in the previous $n$ characters. In
other words LZ77 looks for the largest integer $m$ such that the
string $x_{n+1},...,x_{n+m}$ already appeared in $x_1,...,x_n$.  The
string found is then codified with two numbers: its length $m$ and the
distance from its previous occurrence. If no already encountered
string starts at position $n$ the zipper simply writes the symbol
appearing in that position in the compressed sequence and starts a new
search from position $n+1$.  \\ From the above description it is
intuitive that LZ77 performs better and better as the number of
processed symbols grows. In particular, for infinitely long strings
(emitted by ergodic sources), its performance is ``optimal'', i.e. the
length of the zipped file divided by the length of the original file
tends to $h/\ln 2$~\cite{wyner-ziv}. The convergence to this limit,
however, is extremely slow. Said code rate the average bits per symbol
needed to encode the sequence, it holds:

\be
\mbox{code rate} 
\simeq h_2 + \mathcal{O}\left(\frac{\ln \ln N }{\ln N}\right)
\label{eq:coderate}
\ee

Notwithstanding its limitations, LZ77 can then be seen as a tool for
estimating the entropy of a sequence. However, the knowledge of $h_2$,
though interesting from a theoretical point of view is often scarcely
useful in applications. For practical purposes, on the other hand,
methods able to make {\em comparisons} between strings are often
required. A very common case, for instance, is that in which one has
to classify an unknown sequence with respect to a dataset of known
strings: i.e. one has to decide which known strings is closer (in some
sense) to the unknown string.

In Section II we have introduced the relative entropy and the cross
entropy between two sources. Recently, a method has been proposed
for the estimate of the cross entropy between two strings based on
LZ77~\cite{bcl}. Recalling that the cross entropy $C(A|B)$ between two
strings $A$ and $B$, is given by the entropy per character of $B$ in
the optimal coding for $A$, the idea is that of appending the two
sequences and zipping the resulting file $A+B$. In this way the zipper
``learns'' the $A$ file and, when encounters the $B$ subsequence,
tries to compress it with a coding optimized for $A$. If $B$ is not
too long~\cite{paper-yak,long}, thus preventing LZ77 from learning it
as well, the cross entropy per character can be estimated as:

\be
C({\cal{A}|\cal{B}}) \simeq \frac{L_{A+B}-L_A}{L_B}
\ee

\noindent where $L_X$ is the length of the compressed $X$ sequence.
This method is strictly related to the Ziv-Merhav
algorithm~\cite{ziv-merhav} to estimate the relative entropy between
two individual sequences.


\section{Examples and numerical results} 

\begin{figure}
\vspace{0.7cm}
\begin{center}
\centerline{\psfig{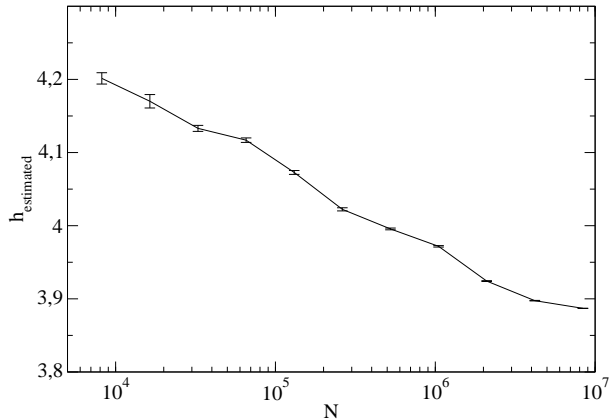}}
\caption{{\bf Entropy estimation:} The number of bits
  per characters of the zipped sequence $h_{estimated}$ is plotted versus
  the length $N$ of the original one. Bernoulli sequences with $K=10$
  symbols are analyzed. The zipper performs better with longer
  strings, but the convergence towards the optimal compression, thought
  theoretically proved, is extremely slow. The Shannon entropy of the
  considered sequences is $h_2 \simeq 3.32$ and, for strings of
  approximately $8 \times 10^6$ characters, $h_{estimated}$ is $18\%$
  larger than this value.}
\label{fig1}
\end{center}
\end{figure}

In this section we illustrate the behavior of LZ77 in experiments of
entropy estimation and of recognition with two examples.
Figure~\ref{fig1} reports the LZ77 code rates when zipping Bernoulli
sequences of various lengths. A Bernoulli string is generated by
extracting randomly one of $K$ allowed symbols with probability $1/K$
($K=10$ in our case). The entropy of such strings is simply $\log
K$. From the Figure it is evident that the zipper performs better and
better with longer strings, though, as seen in~(\ref{eq:coderate}),
the convergence is extremely slow. It is important to remark how there
exist more efficient algorithms to estimate the entropy of a string.
We refer the interested reader to~\cite{grassberger} for a recent
review. It is nevertheless useful to quote the so-called Shannon Game
to estimate the entropy of English (see~\cite{shannongame} for an
applet) where the identification of the entropy with a measure of
surprise is particularly clear.

\begin{figure}
\centerline{\psfig{figure=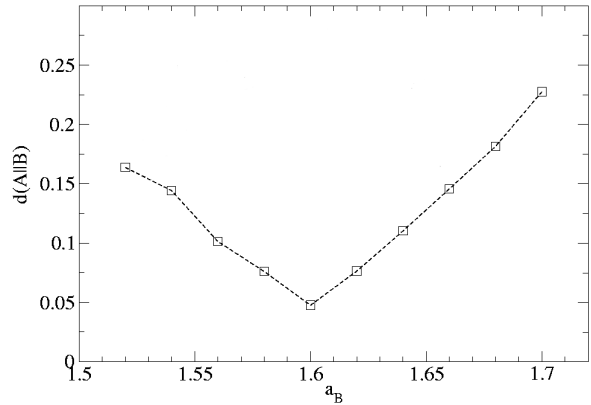,width=8cm,angle=0}}
\caption{{\bf Recognition experiment:} The relative
  entropy, estimated by means of LZ77 as discussed in text, between an
  unknown sequence $A$ and a set of known strings $B$ allows to
  identify the source of $A$. All the sequences are generated by a
  Lozi map, and the problem is to identify the parameter $a_A=1.6$ of
  the source of $A$. The minimum of relative entropy allows clearly to
  identify this parameters indicating that $A$ is closer to the string
  $B$ generated with $a_B=a_A$ than to any strings generated with
  different values of $a$.}
\label{fig2}
\end{figure}

In Figure~\ref{fig2} an experiment of recognition is
reported~\cite{paper-yak}.  Here an unknown sequence is compared, in
the sense discussed in the previous section, with a number of known
strings. The idea to test is that the unknown sequence was emitted by
the same source of the closer known one. The source is here a Lozi
map, i.e. a dynamical system of the form:

\begin{displaymath}
\left\{
\begin{array}{rcl}
x_{n+1} & = & 1 - a |x_n| + y_n \\
y_{n+1} & = & b x_n \nonumber
\end{array}
\right.
\end{displaymath}

\noindent where $a$ and $b$ are parameters. The sequence of symbols
used in the following test is obtained taking $0$ when $x \le 0$ and
$1$ when $x > 0$. For $b=0.5$, numerical studies show that the Lozi
map is chaotic for $a$ in the interval $(1.51,1.7)$. For a discussion
of the Lozi map, computation of Lyapunov exponents and representation
of its symbolic dynamics in terms of Markov chains,
see~\cite{matr_crisanti}.

Figure~\ref{fig2} reports the result of this test. A Lozi map with
$a=1.6$, $b=0.5$ and initial condition $x=0.1$, $y=0.1$ has been used
to generate the sequence $A$, of length $10000$, that will be used as
unknown sequence. As probing sequences we have generated a sets of
sequences, $B$ of length $1000$, obtained with Lozi maps with the
parameters $b=0.5$ and $a_B$ varying between $1.52$ and $1.7$. The
quantity computed and reported in the graph is an estimate of the
Kullback-Leibler entropy $d(B \vert \vert A)=C(A|B)-C(B^*|B)$, where
$C(B^*|B)$ is the estimate, in the framework of our scheme, of the
entropy rate of $B$ and $B^*$ is another set of sequences of length
$10000$. As it is evident, in our experiment, the closer sequence to
the unknown one is the one with $a=1.6$ and this means that the
recognition experiments was successful.


\section{Conclusions}

The possibility of quantifying the information contained in a string
of symbols has been one of the great advancements in science of the
last 60 years. Both Shannon's and Algorithmic approaches indicate that
finding a synthetic description of a string is a way to determine how
much information it stores. It is then natural focusing on those
algorithms conceived expressly to compress a string, also known as
zippers. In this paper we have introduced some fundamental concepts of
Information Theory and we have described the LZ77 compressor. This
zipper has the property of being asymptotically optimal, thus being
also a potential tool for estimating the entropy of a string. More
interestingly, we have discussed the possibility for LZ77 to be used
for the estimation of such quantities such as cross or relative
entropy which measure the remoteness between different strings.
Finally we have shown a simple example of entropy estimation for a
Bernoulli sequence and a successful experiment of recognition between
strings emitted by a Lozi map with different
parameters. 

\section*{Acknowledgments}
we thank Valentina Alfi, Dario Benedetto,
Andrea Puglisi and Angelo Vulpiani for many interesting discussions
and contributions to this work. V.L. acknowledges the partial support
of the ECAgents project funded by the Future and Emerging Technologies
program (IST-FET) of the European Commission under the EU RD contract
IST-1940. E.C. acknowledges the partial support of the European
Commission through its 6th Framework Programme "Structuring the
European Research Area" and the contract Nr. RITA-CT-2004-505493 for
the provision of Transnational Access implemented as Specific Support
Action.

\section*{Appendix}
We report here an example of implementation of LZ77. It must be
intended as a didactic illustration, since actual implementations of
the algorithm contain several optimizations.  {\footnotesize
\begin{itemize}
\item Build a vector $V$ whose $j^{th}$ component $V[j]$ is the  $j^{th}$
  symbol of string $S$ that must be compressed;
\item Build a vector $I$ whose $j^{th}$ component, $I[j]$, is the position of
  the closest previous occurrence of symbol $v$ appearing in $V[j]$, or
  $0$ if symbol $v$ has never appeared before;
\item Build an empty vector $C$ which will contain the processed $V$
  (i.e. the processed $S$);
\item define $i=0$; 

  while ($i<|V|$) do: 

 \hspace{0.7cm} define $p=I[i]$, $lmax=1$, $pmax=0$; 

 \hspace{0.7cm} while ($p\neq0$) do: 

 \hspace{1.4cm} define $l=1$;

 \hspace{1.4cm} while ($V[i+l]=V[p+l]$ and $(p+l)<i $)
 do: 

 \hspace{2.1cm} $l=l+1$; 

 \hspace{1.4cm} if $l>lmax$ do $lmax=l$, $pmax=p$;

 \hspace{1.4cm} $p=I[p]$;

 \hspace{0.7cm} if $(l>1)$ append to vector $C$ the token
 $(lmax,i-pmax)$;

 \hspace{0.7cm} else, if $(l=1)$, append to vector $C$ the token
 $(0,0,V[i])$;

 \hspace{0.7cm} $i=i+l$;
\end{itemize}
}

\noindent Before concluding we mention two of the most common adjoint
features to the LZ77 algorithm. The first aims at codifying better the
length-distance token. This is often achieved by zipping further the
compressed string exploiting its statistical properties with the
Huffman algorithm [23]. The second feature is due to the necessity of
speeding up the zipping process in commercial zippers. It consists in
preventing LZ77 from looking back for more than a certain number $w$
of symbols. Such a modified zipper is said to have a ``w-long sliding
window''.

\end{document}